\documentclass[prd,superscriptaddress,twocolumn,nofootinbib,showpacs,preprintnumbers,floatfix]{revtex4}

\usepackage{mathrsfs}
\usepackage{amsfonts,amssymb,amsmath}
\usepackage{graphicx,epsfig}
\usepackage{multirow}

\begin{document}

\title{Natural Predictions for the Higgs Boson Mass \\ and Supersymmetric Contributions to Rare Processes}

\author{Tianjun Li}

\affiliation{Key Laboratory of Frontiers in Theoretical Physics, Institute of Theoretical Physics,
Chinese Academy of Sciences, Beijing 100190, P. R. China }

\affiliation{George P. and Cynthia W. Mitchell Institute for Fundamental Physics and Astronomy,
Texas A$\&$M University, College Station, TX 77843, USA }

\author{James A. Maxin}

\affiliation{George P. and Cynthia W. Mitchell Institute for Fundamental Physics and Astronomy,
Texas A$\&$M University, College Station, TX 77843, USA }

\author{Dimitri V. Nanopoulos}

\affiliation{George P. and Cynthia W. Mitchell Institute for Fundamental Physics and Astronomy,
Texas A$\&$M University, College Station, TX 77843, USA }

\affiliation{Astroparticle Physics Group, Houston Advanced Research Center (HARC),
Mitchell Campus, Woodlands, TX 77381, USA}

\affiliation{Academy of Athens, Division of Natural Sciences,
28 Panepistimiou Avenue, Athens 10679, Greece }

\author{Joel W. Walker}

\affiliation{Department of Physics, Sam Houston State University,
Huntsville, TX 77341, USA }


\begin{abstract}

\textit{For John Ellis  on the celebration of his 65th birthday...} \\

In the context of No-Scale $\cal{F}$-$SU(5)$, a model defined by the convergence of the ${\cal F}$-lipped $SU(5)$ Grand Unified Theory, two pairs of hypothetical TeV scale vector-like supersymmetric multiplets with origins in ${\cal F}$-theory, and the dynamically established boundary conditions of No-Scale Supergravity, we predict that the lightest CP-even Higgs boson mass lies within the range of 119.0 GeV to 123.5 GeV, exclusive of the vector-like particle contribution to the mass. With reports by the CMS, ATLAS, CDF, and D\O~Collaborations detailing enticing statistical excesses in the vicinity of 120 GeV in searches for the Standard Model Higgs boson, all signs point to an imminent discovery.  While basic supersymmetric constructions such as mSUGRA and the CMSSM have already suffered overwhelming reductions in viable parameterization during the LHC's initial year of operation, about 80\% of the original No-Scale $\cal{F}$-$SU(5)$ model space remains viable after analysis of the first 1.1~${\rm fb}^{-1}$ of integrated luminosity. This model is moreover capable of handily explaining the small excesses recently reported in the CMS multijet supersymmetry search, and also features a highly favorable ``golden'' subspace which may simultaneously account for the key rare process limits on 
the muon anomalous magnetic moment $(g - 2)_\mu$ and the branching ratio of the flavor-changing neutral current decay $b \to s\gamma$.
In addition, the isolated mass parameter responsible for the global particle mass normalization, the gaugino boundary mass $M_{1/2}$, is
dynamically determined at a secondary local minimization of the minimum of the Higgs potential $V_{\rm min}$, in a manner which is deeply
consistent with all precision measurements at the physical electroweak scale.

\end{abstract}


\preprint{ACT-14-11, MIFPA-11-43}

\maketitle


\section{Introduction}

The Large Hadron collider (LHC) has accumulated to date up to 2.3~${\rm fb}^{-1}$ of data from proton-proton collisions at a center-of-mass
beam energy of ${\sqrt s}$ = 7 TeV, already establishing firm constraints on the mass of the lightest CP-even Higgs boson. The CMS~\cite{PAS-HIG-11-022}
and ATLAS~\cite{ATLAS-CONF-135,ATLAS:2011ww} Collaborations have uncovered appealing statistical excesses that hint of the properties of the Standard
Model (SM) Higgs boson, though not yet approaching the five standard deviations essential to claim a conclusive discovery.  CMS has reported a surplus of observed events above the Standard Model background estimation near 120 GeV, positioned directly at a location where background competition against observation is particularly severe.  Nevertheless, the extraordinarily rapid ramping up of the LHC luminosity has allowed large quantities of new data to be sufficiently swiftly amassed that a definitive resolution to the dual questions of the existence and mass of the Higgs boson could be imminent. Moreover, these observations beyond background expectations are also in good agreement with newly established constraints from searches for the Higgs boson by the CDF and D\O~Collaborations~\cite{:2011ra}.  No equally suggestive signal of supersymmetry has thus far been detected by
CMS~\cite{Khachatryan:2011tk,Chatrchyan:2011bz,Chatrchyan:2011bj,Collaboration:2011ida,Chatrchyan:2011ek,Collaboration:2011qs,PAS-SUS-11-003} or
ATLAS~\cite{daCosta:2011hh,daCosta:2011qk,Aad:2011ks,Aad:2011xm,Collaboration:2011jz}, so that one may suspect the LHC's best initial chance to make
a key discovery rests in all probability with the Higgs boson.

The anticipation for discovery of physics beyond the SM at the LHC is fervent, heightening attention on the task of ascertaining what particle physics
models exist which can naturally accommodate, or even perhaps uniquely predict, a Higgs boson in the neighborhood of 120~GeV. The foremost contender for an extension to the SM is Supersymmetry (SUSY), a natural solution to the gauge hierarchy problem. Supersymmetric Grand Unified Theories (GUTs) with gravity mediated supersymmetry breaking, known in their simplest variations as minimal Supergravity (mSUGRA) and the Constrained Minimal Supersymmetric Standard Model (CMSSM), have been exhaustively assessed against the first 1.1~${\rm fb}^{-1}$ of integrated luminosity; an overwhelming majority of the formerly experimentally viable parameter space of these models has failed to survive this testing, and has now fallen out of favor.  This fuels the question of whether there endure SUSY and/or superstring post-Standard Model extensions that can continue to successfully counter the rapidly advancing constraints while simultaneously providing a naturally derived Higgs boson mass near 120 GeV, and while remaining potentially visible to the early operation of the LHC.

An attractive candidate solution to this dilemma may be found in a class of models named No-Scale $\cal{F}$-$SU(5)$~\cite{Li:2010ws, Li:2010mi,Li:2010uu,Li:2011dw, Li:2011hr, Maxin:2011hy, Li:2011xu, Li:2011in,Li:2011gh,Li:2011rp,Li:2011fu,Li:2011ex,Li:2011av}. It has been demonstrated that a majority of the bare-minimally constrained~\cite{Li:2011xu} parameter space of No-Scale $\cal{F}$-$SU(5)$, as defined by consistency with the world average top-quark mass $m_{\rm t}$, the dynamically established boundary conditions of No-Scale supergravity, radiative electroweak symmetry breaking, the centrally observed WMAP7 CDM relic density~\cite{Komatsu:2010fb}, and precision LEP constraints on the lightest CP-even Higgs boson $m_{h}$~\cite{Barate:2003sz,Yao:2006px} and other light SUSY chargino and neutralino mass content, remains viable even after careful comparison against the first 1.1~${\rm fb}^{-1}$~\cite{Li:2011fu,Li:2011av} of LHC data.  We shall show that exclusive of the vector-like particle contribution, the light Higgs mass is stably predicted within this region to take a value between 119.0-123.5 GeV, consistent with the surplus of observed events in the analyses presented by the CMS, CDF, and D\O~Collaborations. Significantly, the most promising subspace of this region includes secondary bounds on the flavor changing neutral current $(b \rightarrow s\gamma)$ process, contributions to the muon anomalous magnetic moment $(g-2)_\mu$, and the rare decay process $B_s^0 \rightarrow \mu^+ \mu^-$,  all of which cohere with spin-independent $\sigma_{SI}$~\cite{Aprile:2011hi} and spin-dependent $\sigma_{SD}$~\cite{Tanaka:2011uf} scattering cross-section bounds on Weakly Interacting Massive Particles (WIMPs), in addition to fresh limits established on the annihilation cross-section $\left\langle \sigma v \right\rangle_{\gamma\gamma}$ of WIMPs using gamma-rays derived from the Fermi Telescope observations~\cite{GeringerSameth:2011iw}\cite{Ackermann2011}. This condensed subspace, an updating of our previously advertised ``Golden Strip''~\cite{Li:2010mi}, offers a more focused prediction of the Higgs mass of around 120-121 GeV.  We emphasize that the prediction of the Higgs in the vicinity of 120~GeV has been an exceedingly natural and robust prediction of No-Scale $\cal{F}$-$SU(5)$, stable across the full model space, which we have consistently advertised over the course of a growing body of work~\cite{Li:2010ws, Li:2010mi,Li:2010uu,Li:2011dw, Li:2011hr, Maxin:2011hy, Li:2011xu, Li:2011in,Li:2011gh,Li:2011rp,Li:2011fu,Li:2011ex,Li:2011av}. The recent embellishments to the experimental support for this standing correlation furnish it with a greatly enhanced immediacy and interest.


\section{The $\cal{F}$-$SU(5)$ Model}


The study launched here is built upon the framework of an explicit model, dubbed No-Scale $\cal{F}$-$SU(5)$~\cite{Li:2010ws, Li:2010mi,Li:2010uu,Li:2011dw, Li:2011hr, Maxin:2011hy, Li:2011xu, Li:2011in,Li:2011gh,Li:2011rp,Li:2011fu,Li:2011ex,Li:2011av}, uniting the ${\cal F}$-lipped $SU(5)$ Grand Unified Theory (GUT)~\cite{Barr:1981qv,Derendinger:1983aj,Antoniadis:1987dx} with two pairs of hypothetical TeV scale vector-like supersymmetric multiplets with origins in ${\cal F}$-theory~\cite{Jiang:2006hf,Jiang:2009zza,Jiang:2009za,Li:2010dp,Li:2010rz} and the dynamically established boundary conditions of No-Scale Supergravity~\cite{Cremmer:1983bf,Ellis:1983sf, Ellis:1983ei, Ellis:1984bm, Lahanas:1986uc}. A more complete review of this model is available in the appendix of Ref.~\cite{Maxin:2011hy}.

Supersymmetry is broken in the hidden sector in the conventional framework, and then its breaking effects are mediated to the observable sector through gravity or gauge interactions. In GUTs with gravity mediated supersymmetry breaking, referred to as minimal supergravity (mSUGRA), the supersymmetry breaking soft terms can be parameterized by four universal parameters: the gaugino mass $M_{1/2}$, scalar mass $M_0$, trilinear soft term A, and the ratio of the low energy Higgs vacuum expectation values (VEVs) tan$\beta$, in addition to the sign of the Higgs bilinear mass term $\mu$. The $\mu$ term and its bilinear soft term $B_{\mu}$ are determined by the Z-boson mass $M_Z$ and tan$\beta$ after electroweak symmetry breaking (EWSB).

In the simplest No-Scale boundary conditions, $M_0$=A=$B_{\mu}$=0, while $M_{1/2}$ may be non-zero at the unification scale, allowing for low energy supersymmetry breaking. This scenario appears to come into its own only when implemented at a scale approaching the Planck mass~\cite{Ellis:2001kg,Ellis:2010jb}. Accordingly, $M_{\cal F}$, the point of the second stage flipped $SU(5)\times U(1)_X$ unification, transpires as a plausible candidate scale only when substantially decoupled from the primary GUT unification of $SU(3)_C\times SU(2)_L$ at the scale $M_{32}$ via a revision to the renormalization group equations (RGE) from the extra ${\cal F}$-Theory vector multiplets~\cite{Li:2010ws,Li:2010mi}. These interdependencies conspire to diminish rather than broaden the level of uncertainty in the model's predicted phenomenology.

Utilizing the dynamically established  boundary conditions of No-Scale Supergravity at the $\cal{F}$-$SU(5)$ unification scale $M_{\cal F}$, we have previously delineated the extraordinarily constrained Golden Point~\cite{Li:2010ws} and aforementioned earliest derived incarnation of the Golden Strip~\cite{Li:2010mi} which satisfied all current experimental constraints while additionally featuring an imminently observable proton decay rate $\tau_p$~\cite{Li:2009fq}.  The most constrictive constraint imposed upon the viable model space is the unification scale boundary on $B_{\mu}$=0. Furthermore, through application of a ``Super No-Scale'' condition for the dynamic stabilization of the stringy modulus related to the $M_{1/2}$ boundary gaugino mass~\cite{Li:2010uu, Li:2011dw,Li:2011xu,Li:2011ex}, this mass along with the ratio of the Higgs VEVs $\tan\beta$~\cite{Li:2010uu,Li:2011dw,Li:2011xu,Li:2011ex} has been dynamically determined.

The complete collection of supersymmetry breaking soft terms evolve from the single parameter $M_{1/2}$ in the simplest No-Scale supergravity, and consequently the particle spectra are proportionally comparable up to an overall rescaling on $M_{1/2}$, leaving the majority of the ``internal'' 
physical properties invariant.  This rescaling capability on $M_{1/2}$ is not generally expected in competing supersymmetry models, due to
the presence of larger parameterization freedom, particularly with respect to a second independent boundary mass $M_0$ for scalar fields. This rescaling symmetry can be broken to a slight degree by the vector-like mass parameter, although the dependence is rather weak.

\begin{figure*}[htp]
	\centering
		\includegraphics[width=0.7\textwidth]{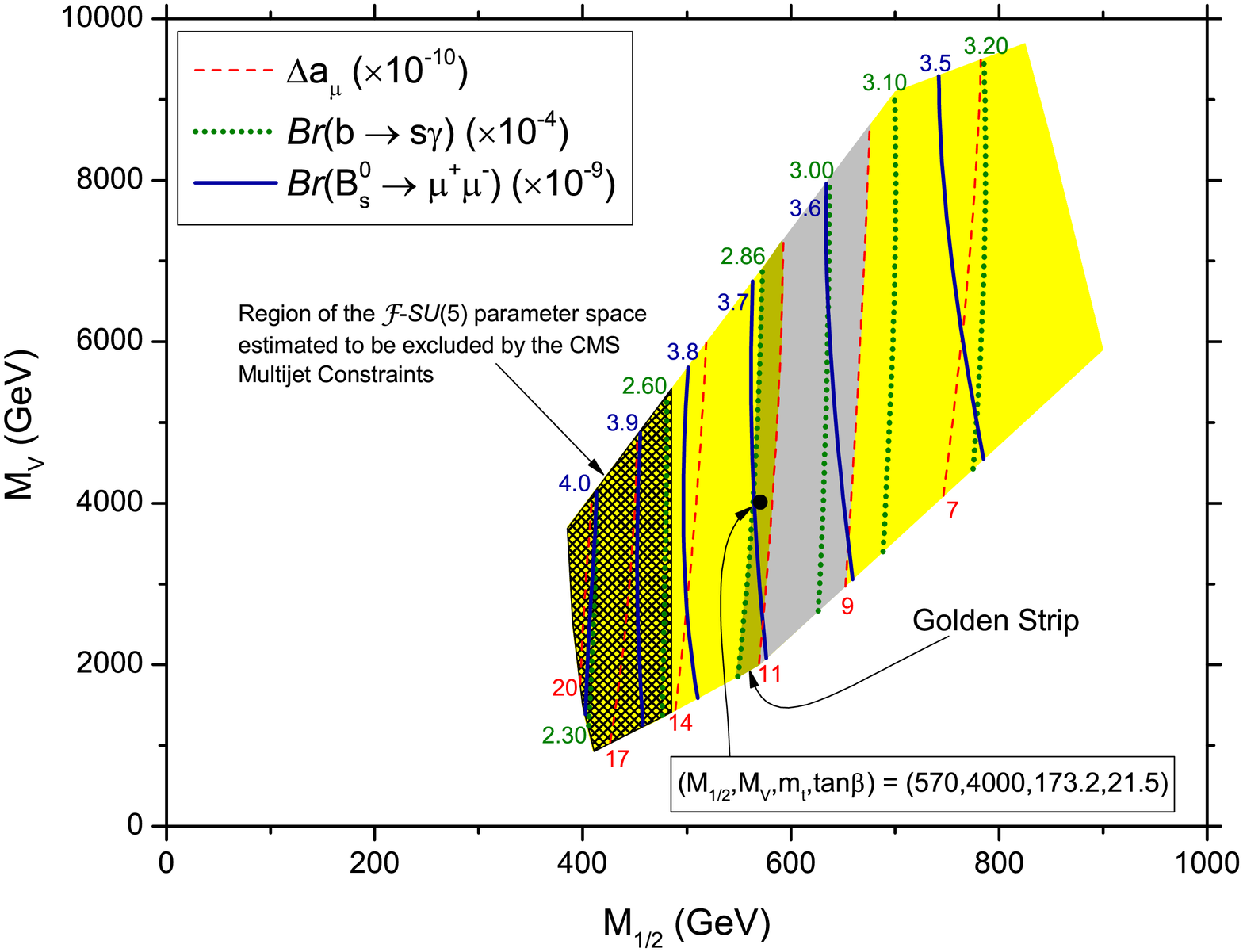}
		\includegraphics[width=0.7\textwidth]{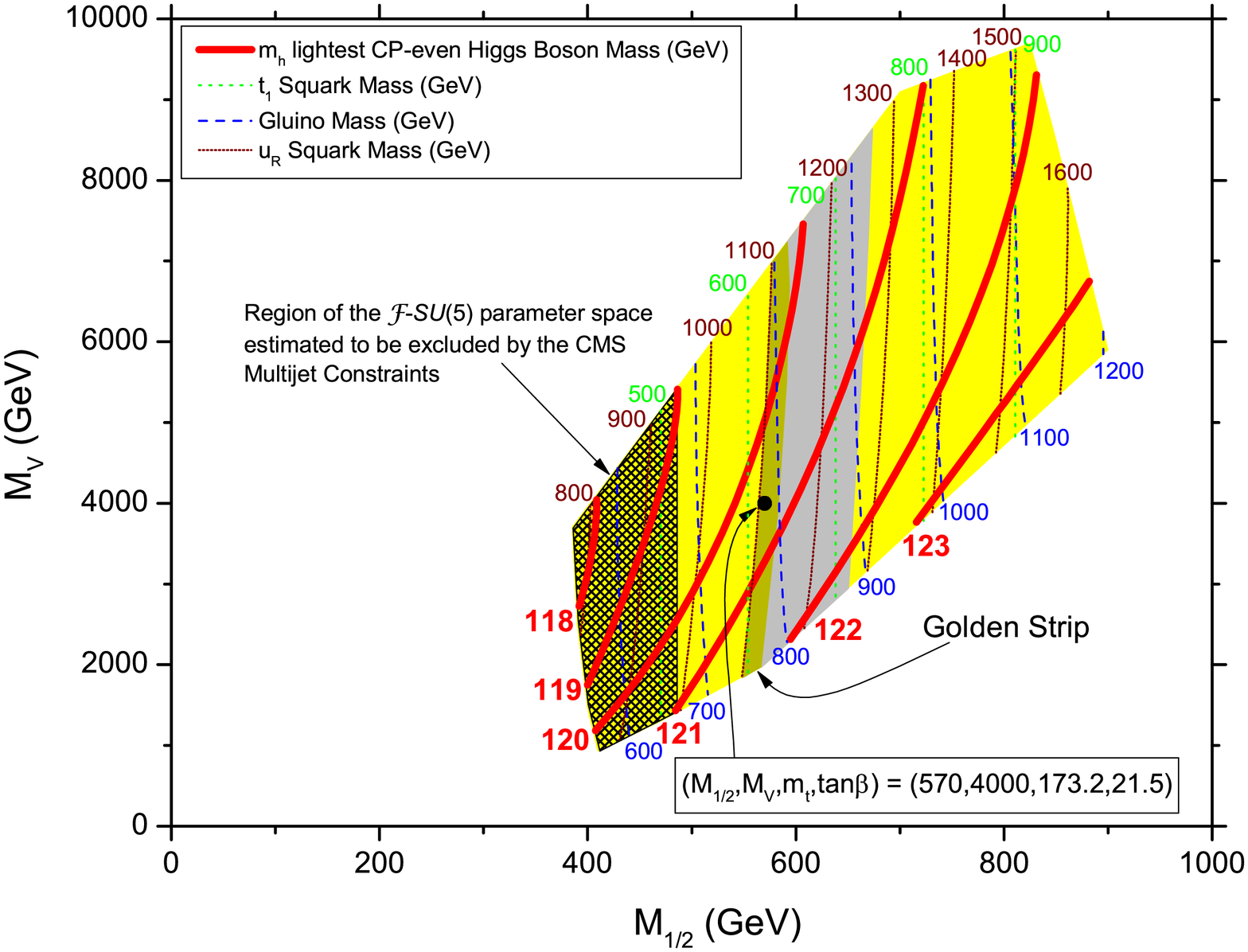}
	\caption{The bare-minimally constrained parameter space of No-Scale $\cal{F}$-$SU(5)$ is depicted as a function of the gaugino boundary mass $M_{1/2}$, the vector-like mass $M_{\rm V}$, and via the solid, dashed, and dotted contour lines, the $(b \rightarrow s\gamma)$, muon anomalous magnetic moment $(g-2)_\mu$, and the $B_s^0 \rightarrow \mu^+ \mu^-$ processes in the upper plot space, with the mass gradients in GeV of the light stop squark $\widetilde{t}_{1}$, gluino $\widetilde{g}$, right-handed up squark $\widetilde{u}_{R}$, and light Higgs mass $m_h$ in the lower plot space. We note that the Higgs mass contours drawn here do not include any additional contributions from the vector-like particles. The region estimated to be disfavored by the first inverse femtobarn of integrated LHC luminosity is marked out with the crosshatch pattern. The vertical strip embossed in gold, referred to as the Golden Strip, represents an experimentally favored region consistent with the bare-minimal experimental constraints of~\cite{Li:2011xu} and both the $(b \rightarrow s\gamma)$ process and contributions to the muon anomalous magnetic moment $(g-2)_\mu$. The Golden Strip also includes the $B_s^0 \rightarrow \mu^+ \mu^-$ decay, however this constraint is satisfied by the entire viable model space. The expanded region adorned in silver imposes these identical constraints, though with a more conservative estimate of $\Delta a_{\mu}=27.5 \pm 18.5 \times 10^{-10}$. The labeled point is the benchmark of Table~(\ref{tab:masses}).}
	\label{fig:Wedge_Masses}
\end{figure*}


\section{Predicting the Higgs Mass}


In the context of No-Scale supergravity, we require $M_0$=A=$B_{\mu}$=0 at the unification scale $M_{\cal F}$, and permit distinctive inputs for the single parameter $M_{1/2}(M_{\cal F})$ to translate under the RGEs to low scale outputs of $B_{\mu}$ and the Higgs mass squares $M_{H_u}^2$ and $M_{H_d}^2$. This evolution continues until the point of spontaneous breakdown of the electroweak symmetry at $M_{H_u}^2$+$\mu^2$=0, at which scale minimization of the broken potential determines the physical low energy values of $\mu$ and tan$\beta$. In practice, we implement this procedure by employing a proprietary modification of the {\tt SuSpect 2.34}~\cite{Djouadi:2002ze} codebase to run the RGEs within {\tt MicrOMEGAs 2.1}~\cite{Belanger:2008sj} to compute the supersymmetry and Higgs mass spectrum. In principle, the $B_{\mu}(M_{\cal F})$=0 condition fixes the value of tan$\beta$ at low energy through running of the RGEs, though this procedure is at odds with the current configuration of the proprietary revised version of the {\tt SuSpect 2.34} codebase. Thus, we equivalently isolate the $B_{\mu}(M_{\cal F})$=0 solution set by instead allowing tan$\beta$ to float freely and subsequently apply a self-consistency check.

For the gauge couplings, we consider two-loop RGE running, though we only consider one-loop RGE running for the gaugino masses, $\mu$ term, supersymmetry breaking scalar masses, trilinear A-terms, and $B_{\mu}$ term. In ${\cal F}$-$SU(5)$ models, the one-loop beta function $b_3$ for $SU(3)_C$ is zero due to the vector-like particle contribution~\cite{Jiang:2006hf}, so the gaugino mass $M_3$ is constant from the electroweak scale to the $M_{32}$ scale~\cite{Li:2010ws}. In contrast, the two-loop gauge coupling RGE running and one-loop gaugino mass RGE running for the $SU(2)_L\times U(1)_Y$ gauge symmetry track each other since the gauge couplings for $SU(2)_L\times U(1)_Y$ are weak; thus, the two loop effects are small. For the calculation of the radiative corrections in the Higgs sector to determine the physical Higgs masses, we implement the full one-loop plus leading two-loop calculations.

To solve the ``$\mu$'' problem for the vector-like particle masses, we can consider the following mechanisms: (1) The Giudice-Masiero mechanism~\cite{Giudice:1988yz}, where such a ``$\mu$'' term is generated from high-dimensional operators; (2) There exist additional Standard Model singlets in F-Theory models. The vector-like particles can couple to these singlets and obtain their masses after these singlets acquire VEVs. This is similar to the solution to the ``$\mu$'' problem in the next-to-the-Minimal Supersymmetric Standard Model (NMSSM).

The vector-like particles can contribute to the Higgs boson masses if they couple to the Higgs fields. For simplicity though, we assume here that such couplings are small. Regardless of this assumption however, our favored benchmark point that we shall introduce in the next section and in Table~(\ref{tab:masses}) possesses an $M_{V}$ of about 4 TeV, hence the contributions of the vector-like particles to the Higgs mass of the benchmark of Table~(\ref{tab:masses}) are small even if their Yukawa couplings to the Higgs fields are not small~\cite{Martin:2009bg}. Though we assume negligible couplings of the vector-like particles to the Higgs fields for the entire model space here in this paper, we shall compute the precise contributions of the vector-like particles to the Higgs boson mass in a forthcoming analysis~\cite{Li:2011ab}.

We take $\mu>0$ as suggested by the results of $(g - 2)_\mu$ for the muon and execute a full scan of the model space through input freedom of the gaugino mass $M_{1/2}$, vector-like mass parameter $M_V$, and tan$\beta$. The resultant solution space is then assessed against the bare-minimal constraint set introduced in Ref.~\cite{Li:2011xu}. To summarize from Ref.~\cite{Li:2011xu}, the bare-minimal constraints are defined by compatibility with the world average top quark mass $m_{\rm t}$ = $173.3\pm 1.1$ GeV~\cite{:1900yx}, the prediction of a suitable candidate source of cold dark matter (CDM) relic density matching the upper and lower thresholds $0.1088 \leq \Omega_{CDM} \leq 0.1158$ set by the WMAP7 measurements~\cite{Komatsu:2010fb}, a rigid prohibition against a charged lightest supersymmetric particle (LSP), conformity with the precision LEP constraints on the lightest CP-even Higgs boson ($m_{h} \geq 114$ GeV~\cite{Barate:2003sz,Yao:2006px}) and other light SUSY chargino, stau, and neutralino mass content, and a self-consistency specification on the dynamically evolved value of $B_\mu$ measured at the boundary scale $M_{\cal{F}}$. An uncertainty of $\pm 1$~GeV on $B_\mu = 0$ is allowed, consistent with the induced variation from fluctuation of the strong coupling within its error bounds and the expected scale of radiative electroweak (EW) corrections.

The cumulative result of the application of the bare-minimal constraints shapes the parameter space into the uniquely formed profile situated in the $M_{1/2},M_{\rm V}$ plane exhibited in Fig.~(\ref{fig:Wedge_Masses}), from a tapered light mass region with a lower bound of $\tan \beta$ = 19.4 into a more expansive heavier region that ceases sharply with the charged stau LSP exclusion around tan$\beta \simeq$ 23. The total model space beyond the hashed over region illustrated in Fig.~(\ref{fig:Wedge_Masses}) consists of those points within the parameter space not excluded by the CMS 1.1~${\rm fb}^{-1}$ constraints, as derived in Ref.~\cite{Li:2011fu}. We demarcate the smooth light Higgs mass $m_h$ gradient in the lower plot space of Fig.~(\ref{fig:Wedge_Masses}) with the emphasized bold contour lines. The region in Fig.~(\ref{fig:Wedge_Masses}) surviving the CMS constraints assertively predicts a quite narrow Higgs mass range of 119.0 to 123.5 GeV, linked to a top quark mass within the world average 173.3 $\pm$1.1 GeV. We note that this range of 119.0 to 123.5 GeV is exclusive of any contributions from the vector-like particles, though we anticipate only a small upward shift in the Higgs mass for the region of the model space with $M_V\gtrsim$ 4 TeV, which includes the benchmark of Table~(\ref{tab:masses}). Nonetheless, this span of Higgs masses thus far agrees well with the excess of data events observed by the CMS~\cite{PAS-HIG-11-022}, CDF and D\O~\cite{:2011ra} Collaborations in the vicinity of 120 GeV. Note also that the Higgs mass in the entire model space is comfortably below the recently derived upper bounds of 145 GeV by CMS~\cite{PAS-HIG-11-022} and 146 GeV by ATLAS~\cite{ATLAS-CONF-135}.


\section{The Golden Strip}


The Golden Strip is strictly defined by the mutual intersection of the bare-minimal constraints with the rare-decay processes $b \to s \gamma$, $B_s^0 \to \mu^+ \mu^-$, and the muon anomalous magnetic moment, as depicted by the condensed vertical slice embossed with gold in both plot spaces of Fig.~(\ref{fig:Wedge_Masses}). For the experimental limits on the flavor changing neutral current process $b \rightarrow s\gamma$, we draw on the two standard deviation limits $Br(b \to s \gamma)=3.52  \pm 0.66 \times 10^{-4}$, where the theoretical and experimental errors are added in quadrature~\cite{Barberio:2007cr, Misiak:2006zs}. We likewise apply the two standard deviation boundaries $\Delta a_{\mu}=27.5 \pm 16.5 \times 10^{-10}$~\cite{Bennett:2004pv} for the anomalous magnetic moment of the muon, $(g - 2)_\mu$. Lastly, we use the recently published upper bound of $Br(B_{s}^{0} \rightarrow \mu^{+}\mu^{-}) < 1.9 \times 10^{-8}$~\cite{Chatrchyan:2011kr} for the process $B_{s}^{0} \rightarrow \mu^+ \mu^-$. The more spacious vertical segment adorned in silver in Fig.~(\ref{fig:Wedge_Masses}) equally consists of all the above constraints, though adopting a more conservative estimate of the $2\sigma$ lower bound of $\Delta a_{\mu} \ge 9.0 \times 10^{-10}$. This shift is supported by a more recent experiment which suggests a downward shift of the central value ~\cite{Hagiwara:2011af}. Moreover, we remark that our greater confidence between these two experimental metrics is with those referencing $b \to s \gamma$, and that since the two key rare process constraints operate in overlapping opposition, the silver region actually comes closer to the central value of this branching ratio. We note that the entire Gold and Silver Strips remain unblemished by the first 1.1~${\rm fb}^{-1}$ of LHC data, representing optimum candidate regions for the discovery of supersymmetry. Additionally, notice that the Higgs mass in the Golden Strip is right about 120 GeV, in accord with the overall combined contributions
of all individual Higgs decay channels observed by CMS above the Standard Model expectations~\cite{PAS-HIG-11-022}.

We select a benchmark from the Golden Strip representing what we believe to be the most optimum point to be assessed against experiment, as identified in Fig.~(\ref{fig:Wedge_Masses}) by the model parameters, with the spectrum of supersymmetric masses given in Table~(\ref{tab:masses}).  At the benchmark, the isolated mass parameter responsible for the global particle mass normalization, namely the gaugino boundary mass $M_{1/2}$=570 GeV, is dynamically determined at a secondary local minimization of the minimum of the Higgs potential $V_{\rm min}$~\cite{Li:2011dw,Li:2011xu,Li:2011ex} in a manner which is deeply consistent with all precision measurements at the physical electroweak scale, and in particular, the Z-boson mass $M_{\rm Z}$ itself~\cite{Li:2011ex}. Supplementing experimental constraints with the dynamical determination of this {\it minimum minimorum} of our universe, this point fulfills the inclusive group of well-established experimental and theoretical constraints, as summarized in Table~(\ref{tab:constraints}), merging a bottom-up experimentally driven analysis with a theoretically motivated top-down approach.

\begin{table}[ht]
  \small
	\centering
	\caption{Spectrum (in GeV) for $M_{1/2}$ = 570 GeV, $M_{V}$ = 4 TeV, $m_{t}$ = 173.2 GeV, tan$\beta$ = 21.5. Here, $\Omega_{\chi}$ = 0.11 and the lightest neutralino is 99.8\% bino.}
		\begin{tabular}{|c|c||c|c||c|c||c|c||c|c||c|c|} \hline		
    $\widetilde{\chi}_{1}^{0}$&$115$&$\widetilde{\chi}_{1}^{\pm}$&$247$&$\widetilde{e}_{R}$&$214$&$\widetilde{t}_{1}$&$623$&$\widetilde{u}_{R}$&$1112$&$m_{h}$&$120.5$\\ \hline
    $\widetilde{\chi}_{2}^{0}$&$247$&$\widetilde{\chi}_{2}^{\pm}$&$925$&$\widetilde{e}_{L}$&$602$&$\widetilde{t}_{2}$&$1039$&$\widetilde{u}_{L}$&$1209$&$m_{A,H}$&$1001$\\ \hline
    $\widetilde{\chi}_{3}^{0}$&$921$&$\widetilde{\nu}_{e/\mu}$&$596$&$\widetilde{\tau}_{1}$&$123$&$\widetilde{b}_{1}$&$995$&$\widetilde{d}_{R}$&$1153$&$m_{H^{\pm}}$&$1004$\\ \hline
    $\widetilde{\chi}_{4}^{0}$&$924$&$\widetilde{\nu}_{\tau}$&$581$&$\widetilde{\tau}_{2}$&$590$&$\widetilde{b}_{2}$&$1101$&$\widetilde{d}_{L}$&$1211$&$\widetilde{g}$&$783$\\ \hline
		\end{tabular}
		\label{tab:masses}
\end{table}


\section{A Smoking Gun Signal}


The intricate evasion of the full company of independent experimental constraints cataloged in the body of Table~(\ref{tab:constraints}) may appear serendipitous, but it is certainly not accidental. The definitive phenomenological signature of No-Scale $\cal{F}$-$SU(5)$ which facilitates this dexterity is the rather unique encoding $M(\widetilde{t}_1) < M(\widetilde{g}) < M(\widetilde{q})$ of the SUSY particle mass hierarchy. This pattern of a stop lightest supersymmetric quark, followed by a gluino which is likewise lighter than the remaining squarks, is stable across the full model space,
and has not been observed to be precisely replicated in any benchmark control sample of the MSSM, and in particular not by any of the ``Snowmass Points and Slopes'' benchmarks~\cite{Allanach:2002nj}. This hierarchy allows No-Scale $\cal{F}$-$SU(5)$ to bypass collider limits on light squark masses much more adroitly than CMSSM constructions with comparably light Lightest Supersymmetric Particles (LSPs). It is moreover directly responsible for a smoking-gun signal of ultra-high ($\ge 9$) jet multiplicity events, which is expected to be prominently visible in LHC searches, given suitable data selection cuts~\cite{Maxin:2011hy,Li:2011hr,Li:2011fu}. The distinctive $\cal{F}$-$SU(5)$ sparticle mass hierarchy responsible for a preponderance of the robust model characteristics summarized in this work is graphically illustrated in the lower plot space of Fig.~(\ref{fig:Wedge_Masses}), where we demarcate the light stop $\widetilde{t}_1$, gluino $\widetilde{g}$, and $\widetilde{u}_R$ squark mass contours.

\begin{table}[ht]
	\centering
	\caption{Conformity with all the measured constraints for the Table~(\ref{tab:masses}) benchmark point $M_{1/2}$ = 570 GeV,
	$M_{V}$ = 4 TeV, $m_{t}$ = 173.2 GeV, tan$\beta$ = 21.5. Here MM is used to designate the {\it minimum minimorum} of our universe.}
		\begin{tabular}{|c|c|} \hline
${\rm Constraint}$&$~~\cal{F}-$SU(5)$~{\rm Value}~~$\\	\hline \hline
$m_h > 114~{\rm GeV}$&$	 120.5~{\rm GeV}$	\\	\hline
$m_t = 173.3 \pm 1.1~{\rm GeV}$&$	173.2~{\rm GeV}$	\\	\hline	
$\Omega_{\widetilde{\chi}_{1}^{0}} = 0.1123 \pm 0.0035$&$	0.1100$	\\	\hline	
$Br(b \to s \gamma)=3.52  \pm 0.66 \times 10^{-4}$&$	2.88 \times 10^{-4}$	\\	\hline	
$\Delta a_{\mu}=27.5 \pm 16.5 \times 10^{-10}$&$	11.5 \times 10^{-10}$	\\	\hline	
$Br(B_s^0 \to \mu^+ \mu^-) \le 1.9 \times 10^{-8}$&$	3.7 \times 10^{-9}$	\\	\hline
$\tau_p \ge 1.0 \times 10^{34} {\rm yr}$&$	5.1 \times 10^{34} {\rm yr}$	\\	\hline
$\sigma_{SI} < 7 \times 10^{-9} {\rm pb}$&$	1.5 \times 10^{-10} {\rm pb}$	\\	\hline
$\sigma_{SD} < 4.5 \times 10^{-3} {\rm pb}$&$	1 \times 10^{-7} {\rm pb}$	\\	\hline
$\left\langle \sigma v \right\rangle_{\gamma\gamma} < 10^{-26} {\rm cm}^3/{\rm s}$&$	2 \times 10^{-28} {\rm cm}^3/{\rm s}$	\\	\hline
$M_{1/2}@M_{Z} = 91.187 \pm 0.001~{\rm GeV}~{\rm MM}$&$	572.2~{\rm GeV}$	\\	\hline
		\end{tabular}
		\label{tab:constraints}
\end{table}

The mechanism of this distinctive signature may be traced to a fact already noted, that the one-loop $\beta$-function $b_3$
of the $SU(3)_C$ gauge symmetry is zero due to the extra vector-like particle contributions~\cite{Jiang:2006hf}.
The effect on the colored gaugino is direct in the running down from the high energy boundary,
leading to the relation $M_3/M_{1/2} \simeq \alpha_3(M_{\rm Z})/\alpha_3(M_{32}) \simeq \mathcal{O}\,(1)$
and precipitating the conspicuously light gluino mass assignment.  The lightness of the stop squark $\widetilde{t}_1$ is
likewise attributed to the large mass splitting expected from the heaviness of the top quark, via its
strong coupling to the Higgs.  The vector-like particles, with a multiplet
structure almost uniquely mandated by avoidance of a Landau pole within the ${\cal F}$-theory model
building~\cite{Jiang:2006hf,Jiang:2009zza,Jiang:2009za,Li:2010dp,Li:2010rz} context, are in turn
necessary in order to achieve a substantial separation between the initial gauge unification of
$SU(3)_C \times SU(2)_{\rm L}$ at $M_{32} \simeq 10^{16}$~GeV, and the secondary unification of
$SU(5)\times U(1)_{\rm X}$ at $M_{\cal F} \simeq 7\times 10^{17}$~GeV.  This elevation of the final
GUT scale, which is possible only within the context of a model with a two-stage
unification like Flipped $SU(5)$, appears likewise to be necessary in order to successfully implement the
No-Scale boundary conditions, and in particular, the vanishing of the Higgs bilinear soft term $B_\mu$.
We emphasize again that this scenario appears to comes into its own only when applied at a unification scale approaching the
Planck mass~\cite{Ellis:2010jb}. The dynamics of No-Scale Supergravity may themselves play an indispensable
role in establishing the cosmological flatness of our Universe, and possibly even in allowing for
the shepherding of a vast multitude of sister universes out of the primordial quantum ``nothingness'',
while maintaining a zero balance of some suitably defined energy function.


\section{LHC Search Strategy}


\begin{figure*}[htp]
	\centering
	\includegraphics[width=0.80\textwidth]{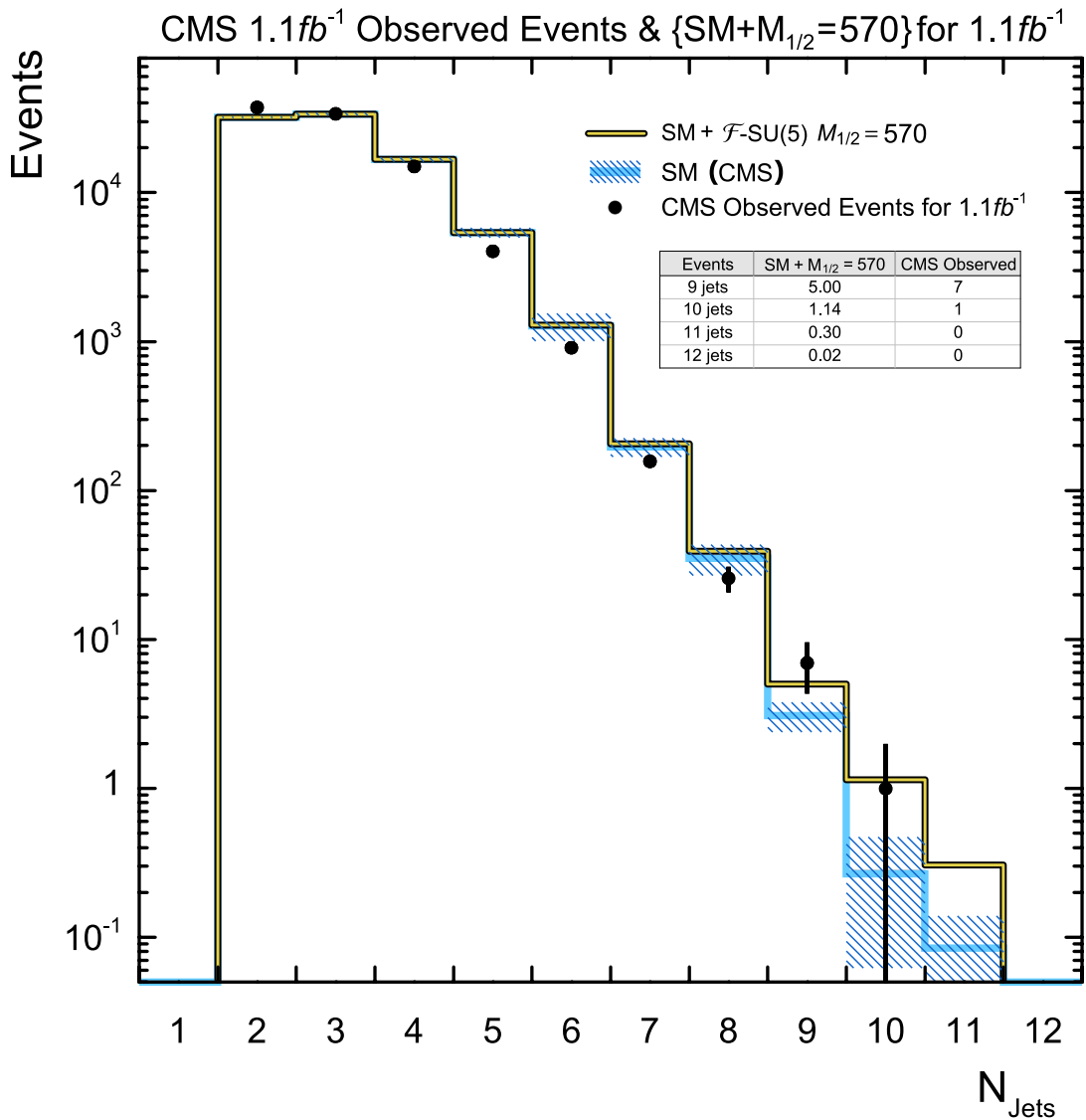}
	\caption{The CMS Preliminary 2011 signal and background statistics for $1.1~{\rm fb}^{-1}$ of integrated luminosity
	at $\sqrt{s} = 7$~TeV, as presented in \cite{PAS-SUS-11-003}, are reprinted with an overlay consisting of a Monte Carlo
	collider-detector simulation of the No-Scale $\cal{F}$-$SU(5)$ model space benchmark of Table~(\ref{tab:masses}). The plot counts events per jet multiplicity, with no cut on $\alpha_{\rm T}$. The Monte Carlo overlay consists of the $\cal{F}$-$SU(5)$ supersymmetry signal plus the Standard Model background, thus permitting a direct visual evaluation against the CMS observed data points.}
	\label{fig:NJets}
\end{figure*}

In Fig.~(\ref{fig:NJets}), we augment the analysis of Ref.~\cite{Li:2011fu} by presenting the number of events generated in a Monte Carlo simulation of our $M_{1/2} = 570$~GeV benchmark point from Table~(\ref{tab:masses}) summed with the Standard Model background statistics of Ref.~\cite{PAS-SUS-11-003}. We superimpose this $\cal{F}$-$SU(5)$ plus Standard Model background signal onto a reprinting of the CMS Preliminary Standard Model background statistics and observed events from Ref.~\cite{PAS-SUS-11-003}, featuring 1.1~${\rm fb}^{-1}$ of collision data and a $\sqrt{s}$ = 7 TeV beam energy. We impose upon the $\cal{F}$-$SU(5)$ signal a set of post-processing cuts designed to mimic those described in the CMS report.  We emphasize that the $\cal{F}$-$SU(5)$ benchmark is quite capable of accounting for the observed event excesses, including most compellingly at the nine jet count, while avoiding any conspicuous overproduction. Although we do here attempt to conform with the $\cal{F}$-$SU(5)$ CMS post-processing cuts presented in Ref.~\cite{PAS-SUS-11-003}, we maintain aggressive advocacy of the ultra-high jet cutting strategy described extensively in Refs.~\cite{Maxin:2011hy,Li:2011hr,Li:2011gh,Li:2011rp,Li:2011fu}. We believe that the discovery of a supersymmetry signal will most likely manifest itself in the data observations for nine or more jets; hence, a jet cutting strategy optimized for extracting supersymmetry from ultra-high jet events could prove to be more efficient at the LHC by one order of magnitude~\cite{Li:2011fu}. 

Furthermore, to emphasize the significance of the ultra-high jet cutting strategy in extracting a No-Scale $\cal{F}$-$SU(5)$ supersymmetry signal, we use the Discovery Index first presented in Ref.~\cite{Li:2011gh} and find that by implementing upon the $M_{1/2}$=570 GeV benchmark point of Table~(\ref{tab:masses}) the CMS post-processing cuts of Ref.~\cite{PAS-SUS-11-003}, though only retaining those events with nine jets or more, requires 8.5$fb^{-1}$ of LHC data in order to achieve a five standard deviation discovery of supersymmetry. With 5 $fb^{-1}$ currently in hand at the LHC as we close the year 2011, and also considering projections that 10$fb^{-1}$ could be attained by the end of the year 2012, a five standard deviation discovery of an $\cal{F}$-$SU(5)$ supersymmetry signal using the CMS search strategy of~\cite{PAS-SUS-11-003} is certainly achievable. However, a prerequisite of utmost importance for the accessibility of such a discovery is that only those events with nine or more jets can be retained. For instance, if all events with 6 or more jets are retained while maintaining the CMS post-processing cutting strategy of~\cite{PAS-SUS-11-003}, then the discovery threshold for $\cal{F}$-$SU(5)$ supersymmetry elevates to about 14$fb^{-1}$. Yet even more grave will be preserving all events with three jets or greater while implementing the CMS cuts of~\cite{Khachatryan:2011tk}, where in this extremely detrimental scenario a massive 100$fb^{-1}$ of luminosity at the LHC will be required for a five standard deviation discovery of an $\cal{F}$-$SU(5)$ supersymmetric signal. Therefore, we would implore the CMS and ATLAS Collaborations to not exclude the examination of events with nine or more jets in the analysis of LHC data; outside of this optimized search region, the supersymmetry signal of models like No-Scale $\cal{F}$-$SU(5)$ will be strongly masked, and possibly undetectable. Given the presently outlined phenomenological attributes that collectively endorse No-Scale
$\cal{F}$-$SU(5)$ as a principal SUSY GUT candidate, diligence in the investigation of its key experimental signature strikes us as rather advisable.

Our simulation was performed using the {\tt MadGraph}~\cite{Stelzer:1994ta,MGME} suite, including the standard
{\tt MadEvent}~\cite{Alwall:2007st}, {\tt PYTHIA}~\cite{Sjostrand:2006za} and {\tt PGS4}~\cite{PGS4} chain, with
post-processing performed by a custom script {\tt CutLHCO}~\cite{cutlhco} (available for download) which executes the desired
cuts, and counts and compiles the associated net statistics.  All 2-body SUSY processes have been included in our simulation,
which follows in all regards the procedure detailed in Ref.~\cite{Maxin:2011hy}. The Monte Carlo is typically oversampled for SUSY processes and scaled down to the requisite luminosity, which can have the effect of suppressing statistical fluctuations.


\section{Conclusions}

While the search for supersymmetry progresses at the LHC with no conclusive signal observed as of this date, the quest for the Higgs boson is rapidly accelerating. All indications from the CMS, ATLAS, CDF, and D\O~Collaborations suggest that a statistically significant observation of the Higgs boson in the vicinity of 120 GeV could be on the near-term horizon, possibly by the end of 2011. It is thus imperative that we begin to spotlight those supersymmetry models capable of engendering a natural prediction for a Higgs boson mass near 120 GeV. We have focused on one such model here by the name of No-Scale $\cal{F}$-$SU(5)$.

Applying only a set of bare-minimal experimental constraints, more than 80\% of the resulting model space of the $\cal{F}$-$SU(5)$ remains viable after the first 1.1~${\rm fb}^{-1}$ of luminosity at the LHC. We found that this entire surviving model space naturally generates a Higgs mass of 119.0-123.5 GeV, in accord with the overall combined contributions of all individual Higgs decay channels observed by CMS above the expected Standard Model background. Though this 119.0-123.5 GeV mass range does not include any contributions from the vector-like particles, we plan to return to this issue in the future to explicitly compute these additional contributions and augment the mass limits as necessary. The benchmark selected for attention in the present work features particularly heavy vector-like multiplets, so that we might for simplicity consider the contribution to the Higgs mass to be suppressed.  We thus anticipate that predictions for this region of the model space will remain stable when future attention is given to higher order effects, although that may not remain strictly true for smaller values of $M_{\rm V}$. Exposing a condensed subspace of this larger region where the bare-minimal constraints intersect the thresholds of the $b \to s \gamma$, $B_s^0 \to \mu^+ \mu^-$, and muon anomalous magnetic moment processes, we have uncovered the most experimentally favorable region, dubbed the Golden Strip, which continues untouched by the rapidly advancing LHC constraints, remaining wholly viable for supersymmetry discovery, while further indicating a Higgs boson at about 120-121 GeV within this favored subspace. Selecting a representative point from a location within the Golden Strip where the dynamical determination of the secondary minimization of the minimum $V_{\rm min}$ of the Higgs potential agrees to high-precision with precision measurements at the electroweak scale, we assessed this benchmark for its ability to fit the CMS multijet data points and elucidate any unexplained statistical excesses in the first 1.1~${\rm fb}^{-1}$ of LHC data reported by the CMS collaboration. The outcome was positive, with an interesting surplus of events at nine jets perfectly explicable within the realm of the No-Scale $\cal{F}$-$SU(5)$ Golden Strip.

For those physicists and non-physicists alike who have been patiently awaiting a categorical discovery of the Higgs boson for decades, the time may be at hand, as an exceedingly plausible prospect of a discovery near 120 GeV looms large over the coming months. Certainly, the first major discovery of the LHC era will generate warranted enthusiasm throughout the high-energy physics community, but we close with a brief suggestion of what the determination of a Higgs boson discovery around 120 GeV might further disclose as to the structure of a more fundamental theory at high energy scales. Given the recent radical curtailing of the dominant mSUGRA and CMSSM model spaces, a Higgs boson near 120 GeV might be interpreted as a rather strongly suggestive piece of evidence to bolster the No-Scale $\cal{F}$-$SU(5)$ framework in particular, and string theory in general.


\begin{acknowledgments}
This research was supported in part 
by the DOE grant DE-FG03-95-Er-40917 (TL and DVN),
by the Natural Science Foundation of China 
under grant numbers 10821504 and 11075194 (TL),
by the Mitchell-Heep Chair in High Energy Physics (JAM),
and by the Sam Houston State University
2011 Enhancement Research Grant program (JWW).
We also thank Sam Houston State University
for providing high performance computing resources.
\end{acknowledgments}


\bibliography{bibliography}

\end{document}